\documentclass[10pt,letterpaper]{article}

\usepackage{opex3}
\usepackage{amsmath,amsfonts}
\usepackage{cite}
\usepackage{bm}
\usepackage[amssymb]{SIunits}
\usepackage{color}

\graphicspath{{Figures/}}

\begin{document}

\title{Planar designs for electromagnetically induced transparency in metamaterials}

\author{Philippe~Tassin,$^{1}$ Lei~Zhang,$^{2}$ Thomas Koschny,$^{2,3}$\\ E.~N.~Economou,$^{3,4}$ and C.~M.~Soukoulis$^{2,3}$}
\address{$^1$Department~of~Applied~Physics~and~Photonics, Vrije~Universiteit~Brussel,\\ Pleinlaan~2, B-1050~Brussel, Belgium}
\email{philippe.tassin@vub.ac.be}
\address{$^2$Ames Laboratory-U.S.~DOE,~and~Department~of~Physics~and~Astronomy,\\ Iowa~State~University, Ames, Iowa~50011, USA}
\address{$^3$Institute~of~Electronic~Structure~and~Lasers~(IESL), FORTH, and~Department~of~Material Science~and~Technology,
University~of~Crete, 71110~Heraklion, Crete, Greece}
\address{$^4$Department of Computational and Data Sciences, George Mason University, Fairfax,\\ Virginia, USA}

\begin{abstract}
We present a planar design of a metamaterial exhibiting electromagnetically induced transparency that is amenable to experimental verification in the microwave frequency band. The design is based on the coupling of a split-ring resonator with a cut-wire in the same plane. We investigate the sensitivity of the parameters of the transmission window on the coupling strength and on the circuit elements of the individual resonators, and we interpret the results in terms of two linearly coupled Lorentzian resonators. Our metamaterial designs combine low losses with the extremely small group velocity associated with the resonant response in the transmission window, rendering them suitable for slow light applications at room temperature.
\end{abstract}

\ocis{(160.3918) Metamaterials; (260.2110) Electromagnetic optics.}


\section{Introduction}\label{Sect:Introduction}
In recent years, there has been considerable interest in metamaterials, i.e., artificial microstructured materials with properties not found with natural materials~\cite{Smith-2004}. If their constituent elements---typically quasi-static electric circuits or plasmonic structures---are significantly smaller than the wavelength of the incident electromagnetic radiation, the material can be treated as an effectively continuous electromagnetic medium~\cite{Smith-2006,Koschny-2005}. With appropriately designed constituents, it is possible to create materials with negative effective permittivity and negative effective permeability in an overlapping frequency band~\cite{Smith-2000,Soukoulis-2006}. These so-called left-handed materials exhibit novel electromagnetic phenomena, such as backward wave propagation, negative refraction, inverse Doppler effect, and radiation tension instead of pressure~\cite{Veselago-1968}. Furthermore, they enable for phase compensation in optical structures, resulting in lenses with subwavelength resolution ~\cite{Pendry-2000} and photonic devices going beyond the diffraction limit~\cite{Engheta-2002,Kockaert-2006,Alu-2007,Tassin-2008}. Through the technique of transformation optics~\cite{Leonhardt-2008}, metamaterials with arbitrary values of the permittivity and permeability ultimately allow for maximal control over light propagation, leading to seemingly exotic applications including invisibility cloaks and perfect optical concentrators~\cite{Leonhardt-2006,Pendry-2006,Rahm-2008}. {Metamaterials are not limited to electromagnetic radiation, but can also be developed for acoustic waves~\cite{Guenneau-2007}.}

The first left-handed material was fabricated in 2001 by combining metal wires exhibiting negative permittivity and split-ring resonators (SRR) exhibiting negative permeability in a single material~\cite{Shelby-2001}. SRRs and closely related structures, which are still commonly used in the microwave band, are now well understood and many studies have been devoted to their optimization~\cite{Katsarakis-2004,Koschny-2004,Garcia-2005,Bilotti-2007}. Due to saturation of the magnetic response of SRRs above terahertz frequencies, other magnetically resonant structures such as slab wire pairs and fishnets were developed for use in the infrared and visible spectra~\cite{Dolling-2006,Shalaev-2007,Soukoulis-2007}. Although direct coupling between constituent elements in metamaterials---especially in the propagation direction in bulk metamaterials---is important for the accurate design of magnetic resonant behavior, only few works have been devoted to the study of such coupling~\cite{Gay-2002,Penciu-2008}.

In this work, we deliberately introduce coupling between two resonant elements in a metamaterial. Under certain conditions, the spectral response of such a coupled structure can be significantly different from the mere superposition of the two independent resonances. Recently, it was suggested that coupled plasmonic resonating elements~\cite{Zhang-2008,Giesen-2008}, coupled SRRs~\cite{Tassin-2008b}, or coupled fish-scale structures~\cite{Zheludev-2008} can exhibit an electromagnetic response that is reminiscent of electromagnetically induced transparency (EIT) in laser-driven atomic systems~\cite{Fleischhauer-2005}. Although the effect described in this paper is fundamentally different from EIT in atomic systems, it is useful to give a short overview of the latter, since we will find that both phenomena have many features in common.

EIT is a coherent process observed in three-level atomic systems, in which two laser beams are resonantly coupled to two energy level transitions with different relaxation times. Due to destructive quantum interference between the amplitudes associated with different excitation pathways, the transmission of a probe laser beam can be greatly enhanced in a narrow frequency band in the middle of the absorption line~\cite{Fleischhauer-2005}. The effect can be most easily explained by the existence of a dark superposition state when both the coupling and the probe laser fields are turned on. This superposition state has vanishing dipole moment, as a result of which it no longer interacts with the laser fields. The absorption spectrum therefore develops a very narrow transmission window in the broader Lorentzian-like absorption peak associated with the transition to the excited state. However, the reduction in absorption length is extremely sensitive to inhomogeneous broadening due to the Doppler effect in atomic vapor samples or size dispersion in quantum dot ensembles. The most remarkable results are therefore obtained by alkali vapors cooled down to liquid helium temperatures.

In this paper, we report designs of planar metamaterial structures in which each element of the periodic structure consists of two coupled resonant circuits: one SRR and one finite wire. In Sec.~\ref{Sect:PlanarEITMM}, we present the electromagnetic response and the constitutive properties of the metamaterial. Subsequently, in Sec.~\ref{Sect:AnalyticalModel}, we give a physical interpretation of the observed EIT effect and we construct an analytical model in terms of the equivalent circuit representation of the coupled wire-SRR. Finally, we investigate in Sec.~\ref{Sect:ParametricStudy} how the absorption and group velocity are influenced by the geometric parameters (resistance, inductance and spatial separation) of the wire and the SRR.

\section{Planar EIT Metamaterials}\label{Sect:PlanarEITMM}
We follow the same idea for a metamaterial exhibiting electromagnetically induced transparency as proposed in Ref.~\cite{Tassin-2008b}, i.e., we use two coupled quasi-static circuits as the constituent element of a metamaterial. One of the circuits is dipole-coupled to the incident circuit (the ``radiative'' resonator), whereas the other circuit has vanishing dipole moment (the ``dark'' resonator). At a certain resonant frequency, the normal modes of the coupled circuit will interfere destructively in the radiative RLC resonator (almost vanishing current), but constructively in the dark RLC resonator (a strong current in the dark circuit loop is excited).

\begin{figure}[b!]
\begin{center}
\includegraphics[width=12cm]{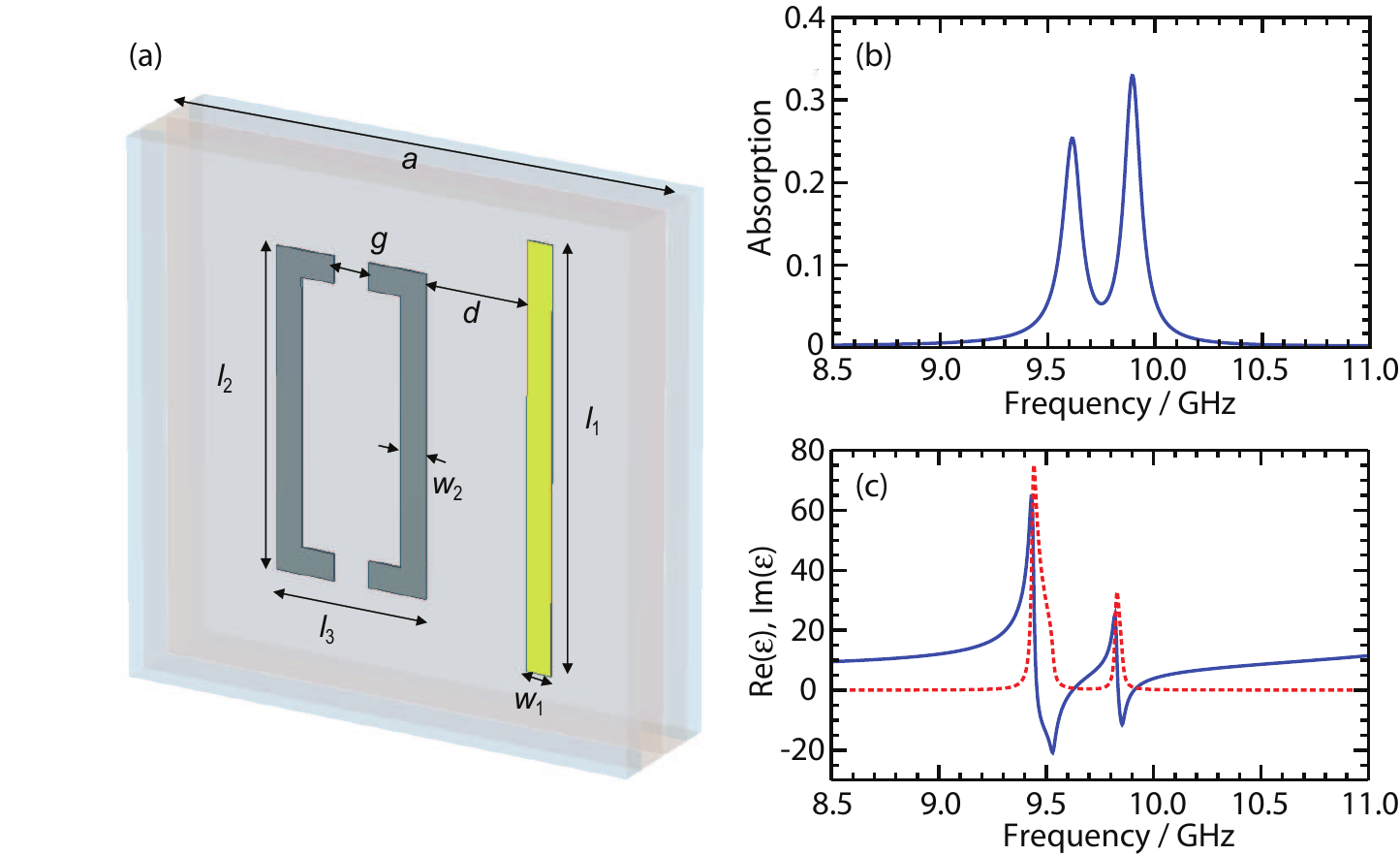}
\caption{(a)~The EIT metamaterial is a square lattice with lattice constant $a = \unit{10}{\milli\meter}$ and has in each unit cell a finite wire capacitively coupled to a (dark) split-ring resonator, supported by a \unit{0.8}{\milli\meter} thick Quartz layer. The cut-wire is made of \unit{35}{\micro\meter} thick copper and has a length of $l_1 = \unit{7.93}{\milli\meter}$ and a width of $w_1 = \unit{0.5}{\milli\meter}$. The two-gap SRR has a total length of $l_2 = \unit{6}{\milli\meter}$ and width of $l_3 = \unit{3}{\milli\meter}$. The wire width of the SRR is $w_2 = \unit{0.5}{\milli\meter}$, the gap size is $g = \unit{0.7}{\milli\meter}$. The separation between the cut-wire and the SRR is $d = \unit{2}{\milli\meter}$. (b)~Absorption spectrum showing the transparency window. (c)~Retrieved permittivity.}
\label{Fig:RefDesign}
\end{center}
\end{figure}

In order to observe an EIT-like effect in the transmission of electromagnetic radiation through this structure, it is crucial that the loss factor of the radiative circuit is significantly larger than the resistance of the dark resonator. In Ref.~\cite{Tassin-2008b}, this was achieved by filling the gaps of the SRRs constituting the dark and radiative circuits with two dielectrics with different loss factor. In this paper, we resort to a different approach: we take different structures for the dark and radiative resonators. For the radiative resonator, we use a cut-wire oriented in the direction of the electric field of the incident electromagnetic wave [see Fig.~\ref{Fig:RefDesign}(a)]. Such a wire will be coupled to the external field by its dipole moment. For the dark resonator, we use a double-split SRR that does not couple to the incident wave by virtue of its symmetry [see Fig.~\ref{Fig:RefDesign}(a)].

We first show that the dark element (the double-split SRR) has the smallest loss factor. To this aim, we remove the wire from the metamaterial, and we calculate the transmission of a plane wave through the metamaterial with only the SRR present by use of a finite integration technique (Microwave Studio). The absorption line of the SRR has a Lorentzian-like shape, from which we can calculate the quality factor by dividing the center frequency $\omega_0$ by the full width half maximum bandwidth, $Q=\omega_0/\Delta\omega_\mathrm{FWHM}$. The same procedure but with the SRR removed is repeated in order to calculate the quality factor of the cut-wire. We find that the quality factor of the dark resonator ($Q_\mathrm{d} \approx 10$) is larger than the quality factor of the radiative circuit ($Q_\mathrm{r} \approx 3.5$), as is necessary in order to observe EIT. The SRR and the cut-wire are also designed such that they have the same resonant frequency at $\omega_0 = \unit{9.35}{\giga\hertz}$.

\begin{figure}[!htb]
\begin{center}
\includegraphics[clip]{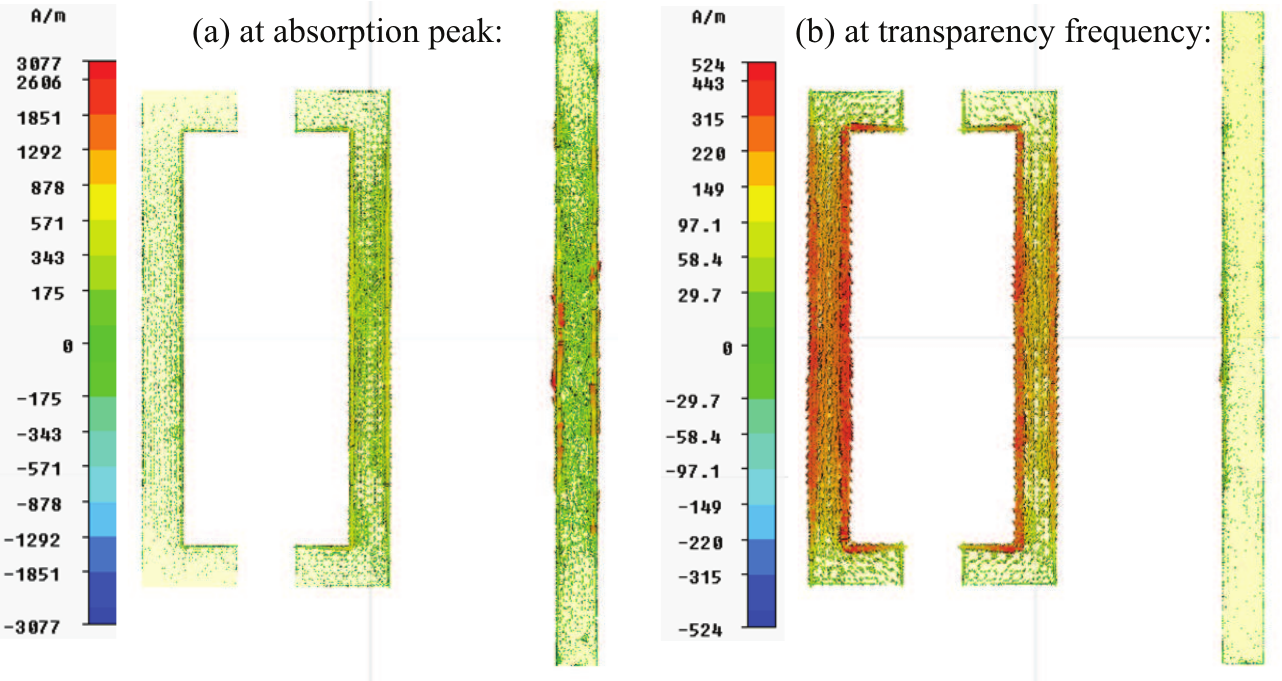}
\caption{Surface current distribution of the wire and the split-ring resonator: (a) at the absorption peak ($f = \unit{9.85}{\giga\hertz}$); (b) at the transparency frequency ($f = \unit{9.72}{\giga\hertz}$).}
\label{Fig:Currents}
\end{center}
\end{figure}

Subsequently, we combine the two resonators and we again calculate the transmission through the metamaterial. It is interesting to have a look at the surface current distributions of the wire and the split-ring resonator at this time (see Fig.~\ref{Fig:Currents}). At the peak absorption ($f = \unit{9.85}{\giga\hertz}$), the highest current can be observed in the wire and the current in the split-ring resonator is significantly smaller; there is only a slight coupling to an electric resonance of the SRR. At the transparency frequency ($f = \unit{9.72}{\giga\hertz}$), the high current in the split-ring resonator shows that the now resonant split-ring is strongly excited; on the other hand, the current in the dipole-coupled wire is small. This behavior is in agreement with our analytical model presented in Sec.~\ref{Sect:ParametricStudy} and also supports our claim that this metamaterial exhibits classical electromagnetically induced transparency.

From the obtained transmission and reflection parameters, we calculate the absorption ($A~=~1 - |S_{11}|^2 - |S_{12}|^2$) and the permittivity {using the parameter retrieval procedure developed by Smith \textit{et al.}~\cite{Smith-2005,Koschny-2005}.} The results are plotted in Fig.~\ref{Fig:RefDesign}(b)-(c). We see that the broad absorption peak, which is mainly due to the dipole-coupled wire, is divided by a narrow transparency region with low absorption and a corresponding small $\mathrm{Im}(\epsilon)$. Inside the transparency window, we observe steep dispersion, which will lead to a significantly increased group index and could be useful for slow-light applications. In the remainder of this paper, we will focus on how we can increase the group index of the proposed metamaterial structure.

\section{Analytical Model for the Metamaterial}\label{Sect:AnalyticalModel}
\begin{figure}[t!]
\begin{center}
\includegraphics[clip]{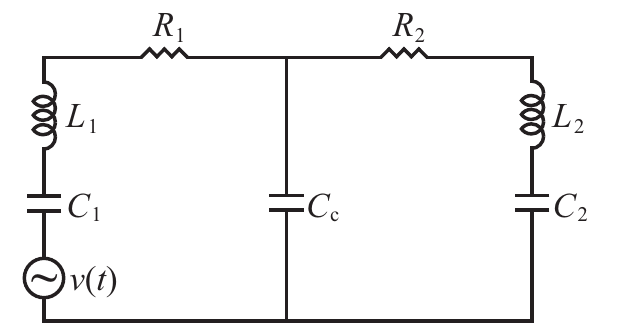}
\caption{Electric circuit modeling the response of the coupled wire-SRR structure.}
\label{Fig:CircuitModel}
\end{center}
\end{figure}

In order to obtain some rules of thumb for the design of an EIT metamaterial with minimal group velocity, we have developed an analytical formulation to calculate how the group index changes as a function of the circuit parameters. We start by modelling the coupled resonators as two RLC circuits coupled by a common capacitor (see Fig.~\ref{Fig:CircuitModel}). The leftmost subcircuit corresponds to the radiative resonator and contains a voltage source representing the action by the external field. The rightmost subcircuit represents the dark circuit and contains no voltage source since it does not interact directly to the external wave.

The currents $i_1$ and $i_2$ circulating in the radiative and dark circuit loops, respectively, can be calculated from standard loop current analysis{~\cite{Caloz-2005}}:
\begin{equation}
\begin{pmatrix}
i_1\\
i_2
\end{pmatrix}
= Z^{-1}\begin{pmatrix}
v\\
0
\end{pmatrix},
\end{equation}
where the impedance matrix $Z$ of the circuit is given by
\begin{equation}
\begin{pmatrix}
-i\omega L_1 + R_1 + \frac{1}{-i\omega C_1} & \frac{1}{-i\omega C_\mathrm{c}}\\
\frac{1}{-i\omega C_\mathrm{c}} & -i\omega L_2 + R_2 + \frac{1}{-i\omega C_2}
\end{pmatrix},
\end{equation}
and we can assume that $v$ is proportional to the local electric field since the cut-wire is a linear system. The dipole moment $p_1$ of the cut-wire will in turn be related to the charge accumulated at its ends, or $p_1 \propto q_1 \propto i_1/(-i\omega)$. Upon application of the Clausius-Mosotti equation to relate the local field to the incident field, we find an analytic formula for the permittivity of the metamaterial:
\begin{equation}
\epsilon = \frac{1+\frac{2}{3}\frac{\beta}{-i\omega}\begin{pmatrix}
1 & 0
\end{pmatrix}Z^{-1}\begin{pmatrix}
1\\
0
\end{pmatrix}}{1-\frac{1}{3}\frac{\beta}{-i\omega}\begin{pmatrix}
1 & 0
\end{pmatrix}Z^{-1}\begin{pmatrix}
1\\
0
\end{pmatrix}},\label{Eq:Epsilon}
\end{equation}
where $\beta$ is proportional to the density of cut-wires in the metamaterial and depends also on a number of unknown proportionality constants that are only function of the exact geometry of the wire. Eq.~(\ref{Eq:Epsilon}) allows to assess quickly how the circuit parameters---and, ultimately, changes in the geometry of the design---affect the properties of the EIT transmission window, such as absorption and group index. Furthermore, since the coupled SRR-wire structure has no magnetic resonances in the EIT transmission window (this is confirmed by the retrieved permeability which is not shown here), the index of refraction can be estimated from $n \approx \sqrt{\epsilon}$ and the group index from $n_g = n + \omega\, \mathrm{d}\,n/\mathrm{d}\omega$.

\begin{figure}[!htb]
\begin{center}
\includegraphics{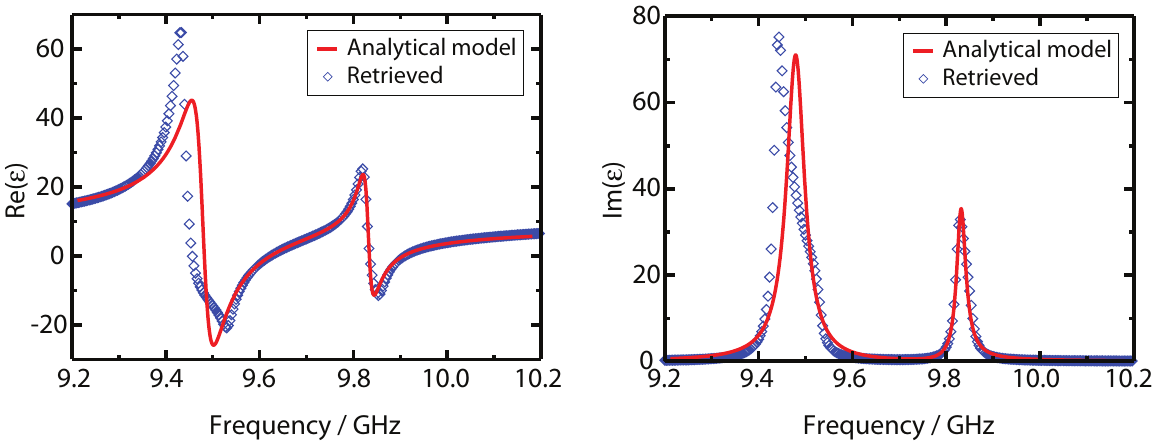}
\caption{Comparison of the analytical model for the permittivity of the metamaterial with the data obtained through the retrieval of the permittivity from the S-parameters obtained by numerical simulation. The red line represents the analytical model of Eq.~(\ref{Eq:Epsilon}); the blue diamonds are the retrieved permittivity data from the numerical simulation.}
\label{Fig:CurveFit}
\end{center}
\end{figure}

We now check the accuracy of the analytical formula by curve fitting Eq.~(\ref{Eq:Epsilon}) to the data obtained for the reference structure depicted in Fig.~\ref{Fig:RefDesign}(a). To obtain the best fit, we use a least squares optimization algorithm. The results are plotted in Fig.~\ref{Fig:CurveFit}. We see that the analytical formula accurately describes the EIT resonance; there are some deviations mainly at the leftmost absorption peak; this is due to either a weaker additional resonance which is of course not described by Eq.~(\ref{Eq:Epsilon}) or due to effects related due to the periodicity of the metamaterial~\cite{Koschny-2005}. Nevertheless, the fit is extremely good in the middle of the transparency window, so it will be adequate for our purpose of estimating the group velocity and absorption around the center frequency.

\section{Influence of Circuit Parameters on the EIT Transmission Window}\label{Sect:ParametricStudy}
In the final section of this paper, we want to study the influence of the circuit parameters on the transmission window. We will focus on the group velocity and the absorption, since these are the two essential parameters for slow light applications {(see Ref.~\cite{Figotin-2006} for a recent review of slow waves photonic media)}. We start again with our analytical formula for the permittivity, Eq.~(\ref{Eq:Epsilon}). Around the center frequency of the transmission window, the current in the wire ($i_1$) is small, so that we can approximate the permittivity by
\begin{equation}
\epsilon \approx 1 - \frac{\beta}{i\omega}\begin{pmatrix}
1 & 0
\end{pmatrix}Z^{-1}\begin{pmatrix}
1\\
0
\end{pmatrix}
\end{equation}
and the group index by
\begin{equation}
n_\mathrm{g} = 1 + \frac{i\beta}{2\omega} \begin{pmatrix}
1 & 0
\end{pmatrix}Z^{-1}\begin{pmatrix}
1\\
0
\end{pmatrix} + \frac{i\beta}{2} \omega \frac{\mathrm{d}}{\mathrm{d}\omega}\frac{1}{\omega}
\left( \begin{pmatrix}
1 & 0
\end{pmatrix}Z^{-1}\begin{pmatrix}
1\\
0
\end{pmatrix}\right).
\end{equation}
If we further simplify this formula by neglecting terms that are small under typical conditions of EIT, we obtain the following formula relating the group index to the circuit parameters:
\begin{equation}
n_\mathrm{g} \approx \beta\omega_0 C_\mathrm{c}^2 L_2,\label{Eq:ApproxGroupIndex}
\end{equation}
with $\omega_0 = 1/(L_1 C_1)^{1/2} = 1/(L_2 C_2)^{1/2}$ the center frequency of the transparency window.

From Eq.~(\ref{Eq:ApproxGroupIndex}), we can observe the following:
\begin{itemize}
\item In first approximation, the group index does not depend on the resistances $R_1$ and $R_2$ and, hence, not on the ohmic losses in the metamaterial.
\item The group index is proportional to $C_\mathrm{c}^2$. This means that weaker coupling (i.e., larger spatial distance between both resonators) leads to smaller group velocities.
\item We have an additional degree of freedom to control the group velocity: the inductance of the dark circuit.
\end{itemize}
We will now illustrate the above statements with numerical simulations. In each of the following examples, we change the geometry of the wire and/or SRR, or the material properties. We then calculate the transmission/reflection (with Microwave Studio), from which we obtain the group index and absorption.

\subsection{Resistance of the radiative resonator}
\begin{figure}[!htb]
\begin{center}
\includegraphics{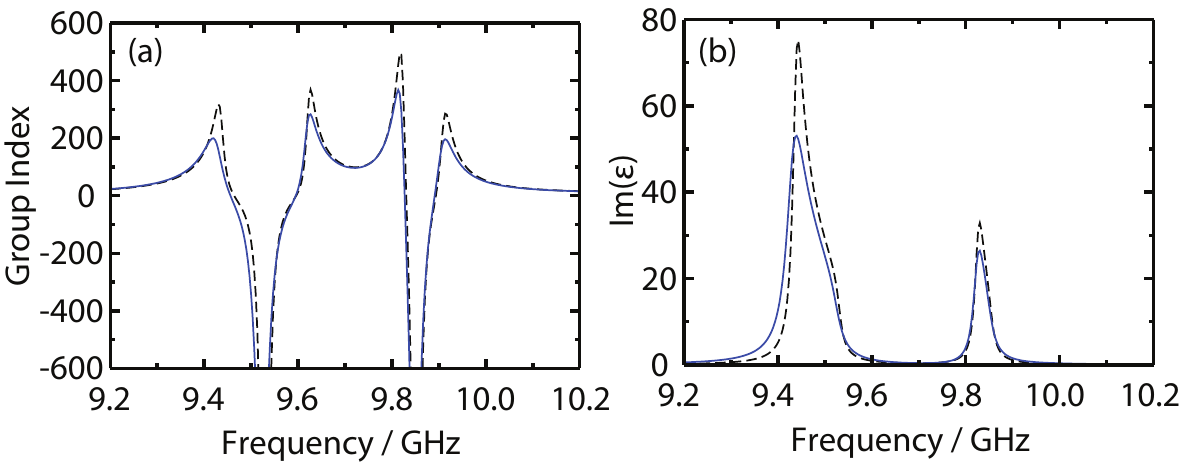}
\caption{(a) Group index, and (b) imaginary part of the permittivity, for the coupled SRR-wire metamaterial. The black dashed lines represent the metamaterial with the same parameters as before; the blue lines represent a metamaterial with higher resistance of the cut-wire.}
\label{Fig:R1}
\end{center}
\end{figure}
We have plotted the group index [Fig.~\ref{Fig:R1}(a)] and the imaginary part of the permittivity [Fig.~\ref{Fig:R1}(b)] for a reference structure with the same parameters as before (black dashed curves) and for a structure with higher $R_1$ (blue curves). {We observe high group index in the frequency range where Fig.~\ref{Fig:RefDesign}(c) has the steepest dispersion.} The resistance of the wire was increased by lowering the conductivity of the metal from $\unit{5.80\times10^7}{\siemens\usk\reciprocal\meter}$ to $\unit{1.16\times10^7}{\siemens\usk\reciprocal\meter}$. We observe that the group index inside the transparency window is almost independent of $R_1$. The absorption is reduced at the peaks, but is more or less unchanged inside the frequency region of interest. Consequently, the resistance $R_1$ is a parameter that hardly influences the properties of the transparency window.

We can easily interpret this behavior. We have seen before that the strong current excited in the dark SRR at frequencies around the transparency frequency dominates the response of the metamaterial. This current does not flow through the resistance $R_1$, and therefore the group velocity and absorption in the transparency window are largely insensitive on the value of $R_1$ or on the losses in the wire.  

\begin{figure}[!htb]
\begin{center}
\includegraphics{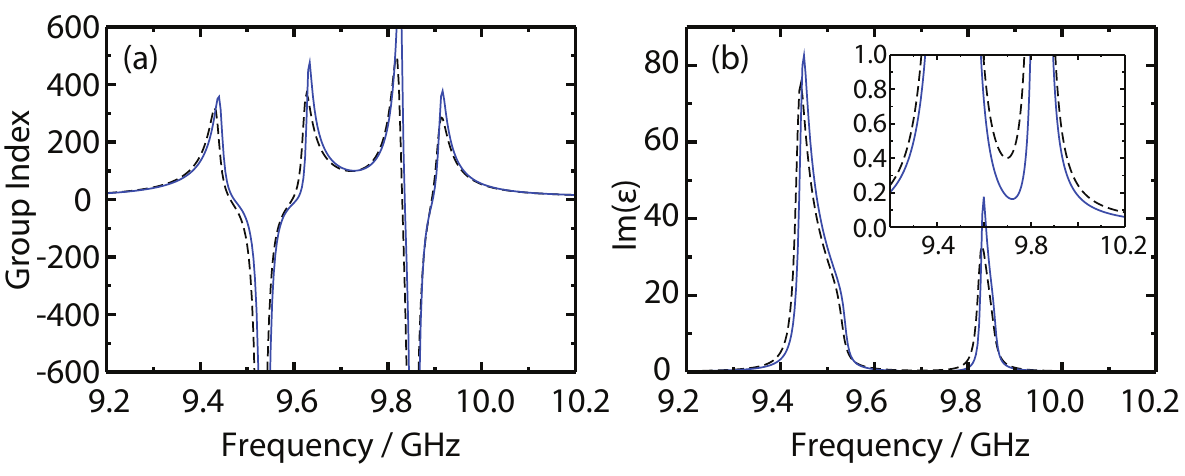}
\caption{(a) Group index, and (b) imaginary part of the permittivity, for the coupled SRR-wire metamaterial. The black dashed lines represent the metamaterial with the same parameters as before; the blue lines represent a metamaterial with smaller resistance of the SRR.}
\label{Fig:R2}
\end{center}
\end{figure}

\subsection{Resistance of the dark resonator}

We again plot the group velocity [Fig.~\ref{Fig:R2}(a)] and the imaginary part of the permittivity [Fig.~\ref{Fig:R2}(b)] for the reference metamaterial (black dashed lines), together with a metamaterial in which we have reduced the resistance of the dark resonator $R_2$ (blue lines). This was achieved by increasing the conductivity of the metal of the SRR with one order of magnitude to $\unit{5.80\times10^8}{\siemens\usk\reciprocal\meter}$. We conclude that the group velocity is only slightly influenced by the resistance of the SRR. However, from the inset of Fig.~\ref{Fig:R2}(b) we can see that this time the absorption inside the transparency window has also decreased. It will therefore be advantageous to minimize the losses in the dark resonator in order to achieve low absorption, which is also an important parameter for slow light applications.

\subsection{Coupling strength}\label{Sec:CouplingStrength}

\begin{figure}[!htb]
\begin{center}
\includegraphics{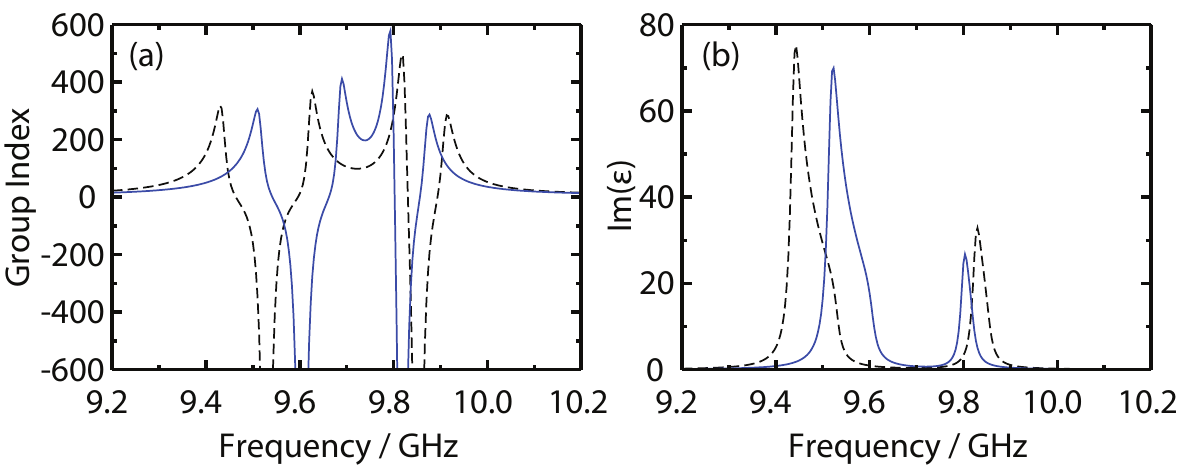}
\caption{(a) Group index, and (b) imaginary part of the permittivity, for the coupled SRR-wire metamaterial. The black dashed lines represent the metamaterial with the same parameters as before; the blue lines represent a metamaterial with weaker coupling, which was obtained by increasing the spatial separation between wire and SRR.}
\label{Fig:Cc}
\end{center}
\end{figure}

From Fig.~\ref{Fig:Cc}(a), we see that the value of the capacitor common to the dark and radiative circuits strongly alters the group index. This parameter can be tuned by changing the spatial separation between the wire and the SRR; in Figs.~\ref{Fig:Cc}(a)-(b), we have have increased the spatial separation from $d = \unit{2.0}{\milli\meter}$ (black dashed lines) to $d = \unit{2.5}{\milli\meter}$ (blue lines). This behavior can be understood by considering the EIT effect as a frequency splitting of the originally degenerate resonances of wire and SRR. The farther away these two elements, the weaker the coupling, and the narrower the transparency window; this in turn leads to stronger dispersion and higher group index. Two remarks must be made here. (i)~The group velocity will be ultimately limited by the size of the unit cell. If the distance between the SRR and the wire is approximately half a lattice constant, the interaction between the wire of one unit cell and the SRR of its nearest neighbor will contribute significantly to the response. Increasing the lattice constant will not help, since the parameter $\beta$ is proportional to the density of the unit cells. (ii)~The increased group velocity goes together with a smaller bandwidth.

\subsection{Inductances of dark/radiative resonator}
\begin{figure}[!htb]
\begin{center}
\includegraphics{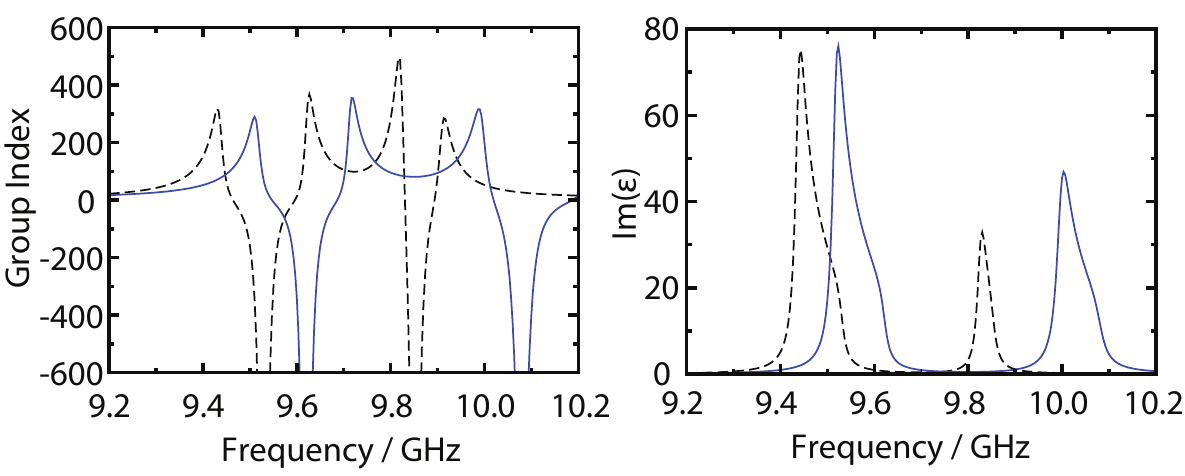}
\caption{(a) Group index, and (b) imaginary part of the permittivity, for the coupled SRR-wire metamaterial. The black dashed lines represent the metamaterial with the same parameters as before; the blue lines represent a metamaterial with smaller inductance in the dark SRR.}
\label{Fig:L2}
\end{center}
\end{figure}

Finally, we investigate the influence of the inductance $L_1$. Our analytical model predicts that a higher inductance of the SRR will lead to a higher group index. This case is somewhat more difficult to check, because a change in the geometry of the SRR will automatically go together with a change in coupling strength. We have therefore first decreased the enclosed surface of the SRR; this will make $L_2$ smaller; at the same time we have also adjusted the thickness of the gaps of the SRRs in order to keep the resonance frequency constant. As to leave the coupling strength unchanged, we have increased the distance between the wire and the SRR until the electric field integrated along the inner loop of the SRR is the same as in the reference case. We can indeed take the integrated electric field as a measure for the coupling strength, since the electric fields in the SRR will increase if the coupling would have been stronger and vice versa.

The results are plotted in Fig.~\ref{Fig:L2}. We see that smaller inductance of the dark resonator, i.e., smaller $L_2$, indeed leads to a somewhat smaller group velocity in the transparency window, although the effect is not very strong. The influence on the absorption is negligible. The first remark made in Sec.~\ref{Sec:CouplingStrength} is also appropriate here, namely that the increase in inductance will finally be limited by the size of the unit cell. On the other hand, it is important to observe that a smaller inductance $L_2$ allows for a significantly larger bandwidth and, hence, delay-bandwidth product, which is another important figure of merit for slow light applications.

\section{Summary}

In this paper, we have demonstrated a planar metamaterial exhibiting an effect reminiscent of electromagnetically induced transparency. The metamaterial consists of a square lattice with a coupled wire-SRR in each unit cell; the use of two different types of metamaterial elements allows for the required difference in quality factor between the two elements. The spectral response of the metamaterial's permittivity features a narrow transparency window inside a much broader absorption peak. We have shown that a strong current is excited in the SRR, even though the SRR is not directly coupled to the incident electromagnetic field. Finally, we have investigated how the different geometrical parameters (characterized by the effective circuit parameters of the dark and radiative resonators) change the properties of the transparency window. The (capacitive) coupling strength between the wire and the SRR is the most important parameter to increase the group index. For the absorption, the resistance of the SRR is crucial. Finally, we found that the inductance of the SRR is also an important parameter with which the bandwidth of the EIT window can be improved.

\section*{Acknowledgments}
P.~T.~is a Research Assistant (Aspirant) of the Research Foundation-Flanders (FWO-Vlaanderen). Work at the Vrije Universiteit Brussel (VUB) was financially supported by the FWO-Vlaanderen, by the university's Research Council (OZR), and by the photonics@be project funded by the Belgian Science Policy Office of the federal government (Contract No.~IAP6/10). Work at Ames Laboratory was supported by the Department of Energy (Basic Energy Sciences) under Contract No.~DE-AC02-07CH11358.  This work was partially supported by the Office of Naval Research (Award No.~N00014-07-1-D359) and European Community FET project PHOME (Contract No.~213390).
\end{document}